\documentclass[journal=jacsat,manuscript=article]{achemso}
\usepackage[version=3]{mhchem} 

\usepackage{graphicx}
\usepackage{dcolumn}
\usepackage{bm}
\usepackage{multirow}
\usepackage[utf8]{inputenc}
\usepackage[T1]{fontenc}
\usepackage{mathptmx}
\usepackage{blindtext}
\usepackage{subcaption}

\usepackage{braket}
\usepackage{graphicx}
\usepackage{amsmath,amssymb,mathtools}
\usepackage{dcolumn}
\usepackage{bm}
\usepackage{color}

\title[]{Quantum-Classical Hybrid Algorithm for the Simulation of All-Electron Correlation}

\author{Jan-Niklas Boyn}
\author{Aleksandr O. Lykhin}
\author{Scott E. Smart}
\author{Laura Gagliardi}
\email{lgagliardi@uchicago.edu}
\author{David A. Mazziotti}
\email{damazz@uchicago.edu}
\affiliation{Department of Chemistry and The James Franck Institute, The University of Chicago, Chicago, IL 60637}

\begin{document}

\begin{tocentry}
\centering
\includegraphics[]{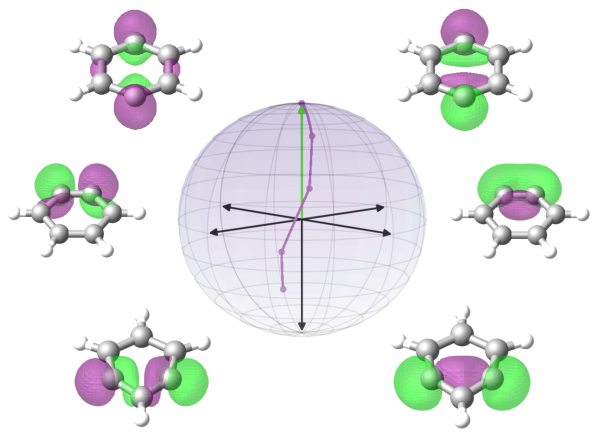}
\end{tocentry}

\begin{abstract}
While the treatment of chemically relevant systems containing hundreds or even thousands of electrons remains beyond the reach of quantum devices, the development of quantum-classical hybrid algorithms to resolve electronic correlation presents a promising pathway toward a quantum advantage in the computation of molecular electronic structure. Such hybrid algorithms treat the exponentially scaling part of the calculation---the static (multireference) correlation---on the quantum computer and the non-exponentially scaling part---the dynamic correlation---on the classical computer.  While a variety of such algorithms have been proposed, due to the dependence on the wave function of most classical methods for dynamic correlation, the development of easy-to-use classical post-processing implementations has been limited.  Here we present a novel hybrid-classical algorithm that computes a molecule's all-electron energy and properties on the classical computer from a critically important simulation of the static correlation on the quantum computer.  Significantly, for the all-electron calculations we circumvent the wave function by using density-matrix methods that only require input of the statically correlated two-electron reduced density matrix (2-RDM), which can be efficiently measured in the quantum simulation.  Although the algorithm is completely general, we test it with two classical 2-RDM methods, the anti-Hermitian contracted Schr\"odinger equation (ACSE) theory and multiconfiguration pair-density functional theory (MC-PDFT), using the recently developed quantum ACSE method for the simulation of the statically correlated 2-RDM.  We obtain experimental accuracy for the relative energies of all three benzyne isomers and thereby, demonstrate the ability of the quantum-classical hybrid algorithms to achieve chemically relevant results and accuracy on currently available quantum computers.
\end{abstract}

\maketitle

\section{Introduction}

Since the advent of density functional theory (DFT)\cite{KSDFT}, electronic structure theory has assumed an evermore important role in chemical research from helping researchers rationalize experiment to guiding design in areas ranging from molecular synthesis to new materials. While research over the last decades has made incredible progress in developing accurate and black-box methods available for use by both theoreticians and experimentalists, the accurate as well as computationally tractable treatment of correlated quantum systems continues to pose a major challenge\cite{KadeRev}. Quantum simulation of molecular systems offers a novel approach to the problem where a strongly correlated wave function can potentially be prepared and measured at non-exponential cost\cite{Reiher_QCRxnMech}.

In this Article we present a novel hybrid-classical algorithm that treats the exponentially scaling part of the calculation---the static (multireference) correlation---on the quantum computer and the non-exponentially part---the dynamic correlation---on the classical computer.  While such methods that treat the multireference and dynamic correlation in two separate calculations are common in classical electronic structure\cite{KadeRev, MCPT, QMC, ChanCanonicalT, ChanDMRG, MRCC, MRPT2}, their adaptation to the quantum computer is non-trivial because most techniques require the multireference-correlation part of the calculation to generate a wave function.  The measurement of such a wave function on the quantum computer, however, scales exponentially with molecular size\cite{Reiher_QCRxnMech, elfving2020quantum}.  As in most molecular simulations, we avoid the exponential scaling by performing a tomography of only the two-electron reduced density matrix (2-RDM)---more specifically, only the statically correlated part of the 2-RDM\cite{qACSE}.  However, in contrast to other methods, we here use the statically correlated 2-RDM as a kernel in two classical correlation methods to generate a system-wide correlated 2-RDM, spanning all of the electrons and orbitals in the calculation.  Importantly, the correlated 2-RDM recovers the all-electron correlation energy and properties of a molecule, thereby enabling larger basis sets and realistic comparisons with experimental results. While other hybrid algorithms use a ``perturb-then-diagnalonize'' strategy to add some dynamic correlation to the Hamiltonian before simulation, most algorithms have avoided the conventional ``diagonalize-then-perturb'' strategy because of the wave function bottleneck\cite{AspuruGuzikRev, KadeRev, ChanRev, elfving2020quantum, Reiher_QCRxnMech, rubin2016hybrid, Bauer_Hybrid, yamazaki2018practical, DeJong_DownfoldingUCC, Motta_Transcorrelated, Takeshita_postQC}.  Although the algorithm is completely general for any 2-RDM-like methods, we test it here with two classical correlation methods: ({\em i}) the anti-Hermitian contracted Schr\"odinger equation (ACSE) theory in which total correlation is computed from a functional of the 2-RDM that is seeded with the statically correlated 2-RDM from the quantum computer\cite{ACSE} and ({\em ii}) multi-configuration pair-density functional theory (MC-PDFT) in which the total correlation is computed as a functional of the statically correlated pair density from the quantum computer\cite{MCPDFT1}.  

We apply the algorithm to resolving the relative energies of the three benzyne isomers to experimental accuracy. The quantum-simulation part of the calculation is performed with the recently developed quantum ACSE (QACSE) algorithm in which a two-electron contraction of the Schr{\"o}dinger equation is solved for the 2-RDM\cite{qACSE, qACSEBenzyne}.  Importantly, because the QACSE algorithm removes either some or all of the classical algorithm's approximation of the higher-order density matrices, it can efficiently capture the static correlation.  Furthermore, the 2-RDMs obtained by solving the QACSE are additionally corrected through an error mitigation scheme that uses $N$-representability conditions. These constraints ensure that a 2-RDM represents at least one $N$-electron density matrix\cite{ScottError}.  The resulting quantum-classical hybrid algorithm accurately captures the energies and properties of the benzynes including the experimental relative energy of the biradical {\em para}-benzyne.     

\section{Results \& Discussion}
\begin{figure}
    \centering
    \includegraphics[scale=0.125]{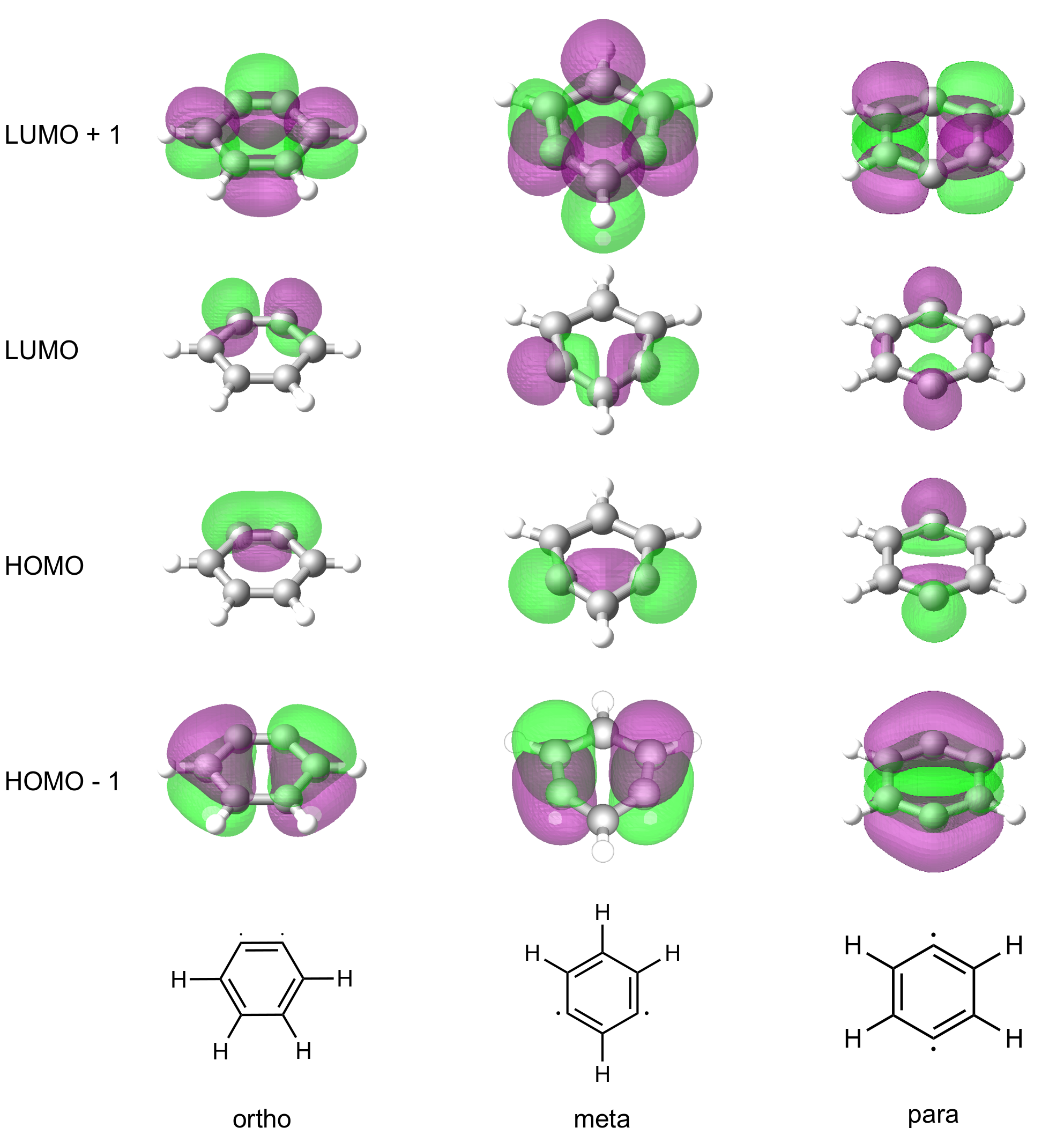}
    \caption{HOMO and LUMO for the three structural benzyne isomers as obtained from classical CASSCF calculations in a [4,4] active space with a cc-pVDZ basis set. The biradical character displayed in the frontier MOs increases as the orbital separation between the radical carbon atoms increases. While spatial proximity allows formation of essentially bonding and anti-bonding orbitals in ortho- and meta-benzyne, yielding near-closed-shell character, the larger distance in the para-isomer yields an electronic structure with strong biradical character. }
    \label{fig:MO_Diagrams}
\end{figure}

The ACSE and its quantum device implementation rely on a contracted form of the Schr\"odinger equation, the CSE\cite{CSE}, which arises from the fact that fermions interact pairwise. Selection of only the anti-Hermitian part of the CSE yields the ACSE:
\begin{equation}
    \bra{\Psi}[\hat{a}^{\dagger}_i \hat{a}^{\dagger}_j \hat{a}_l  \hat{a}_k,\hat{H}]\ket{\Psi} = 0 \,.
\end{equation}
This equation may be solved via a series of differential equations, either on a classical computer\cite{ACSE2, ACSE3} or using a recently introduced quantum device implementation as a contracted quantum eigensolver, termed QACSE\cite{qACSE}. While the classical implementation of the ACSE resolves dynamic correlation with comparable accuracy to CCSD(T)\cite{ACSECC}, it may be seeded with a correlated guess 2-RDM from a CAS calculation to resolve the total correlation energy within a specified active space \cite{ACSE2, ACSE3}.  The QACSE is able to resolve static and other strong correlation accurately because in contrast to the classical algorithms, it does not require approximate reconstruction of the three-electron RDM (3-RDM) from the 2-RDM. Effective error correction on NISQ devices is achieved via the enforcement of $N$-representability conditions on the measured 2-RDM\cite{ScottError, Nrep}. \\

MC-PDFT in turn is an extension of DFT to strongly correlated systems. Seeded with a strongly correlated 2-RDM obtained from a CAS calculation, kinetic and classical Coulomb energies are evaluated and the rest of the energy is non-iteratively computed as a functional of the total electron density, the on-top pair density, and their spatial derivatives using one of the various available functionals. In this article we survey translated\cite{MCPDFT1} (tPBE and tBLYP), (ii) fully-translated\cite{ftPBE} (ftPBE), and (iii) translated hybrid\cite{tPBE0} (tPBE0) density functionals. The ACSE calculations were performed with the Maple Quantum Chemistry Package\cite{Maple, QCP} and MC-PDFT calculations were performed in the PySCF package\cite{PySCF} augmented with the {\em mrh} addon\cite{mrh} that enables MC-PDFT features.

As a benchmarking case of correlated organic molecules, we consider the three structural isomers of benzyne\cite{Benzyne1, Benzyne2, BenzyneExp}. These are obtained by the elimination of two substituents on different positions of the benzene ring, giving rise to ortho-, meta- and para- isomers. While all benzyne isomers are usually drawn as biradical structures, the degree to which they display multi-reference character varies according to the spatial separation between the radical electrons' location on the benzene ring. Figure~\ref{fig:MO_Diagrams} displays the frontier molecular orbitals for each of the three structural benzyne isomers, showing the clear variation in bonding between the ortho, meta, and para geometries. Spatial proximity between the two radical carbon atoms on the benzene ring allows substantial bond formation in the ortho and meta cases, leading to near-closed-shell character, while the para geometry, characterized by large spatial separation between the radical carbons, allows for no overlap between the carbon atomic orbitals, resulting in strong biradical character. The stability of the three isomers correlates with the proximity of the radical carbon atoms, with experimental measurements yielding energies of $15.3 \pm 4.31$ kcal/mol and $31.2 \pm 4.17$ kcal/mol, for the meta and para isomers, relative to the ortho structure, respectively\cite{BenzyneExp}.  \\

\subsection{Relative Energies}
\begin{figure*}
    \centering
    \begin{subfigure}[t]{.48\textwidth}
        \centering
        \includegraphics[scale=0.275]{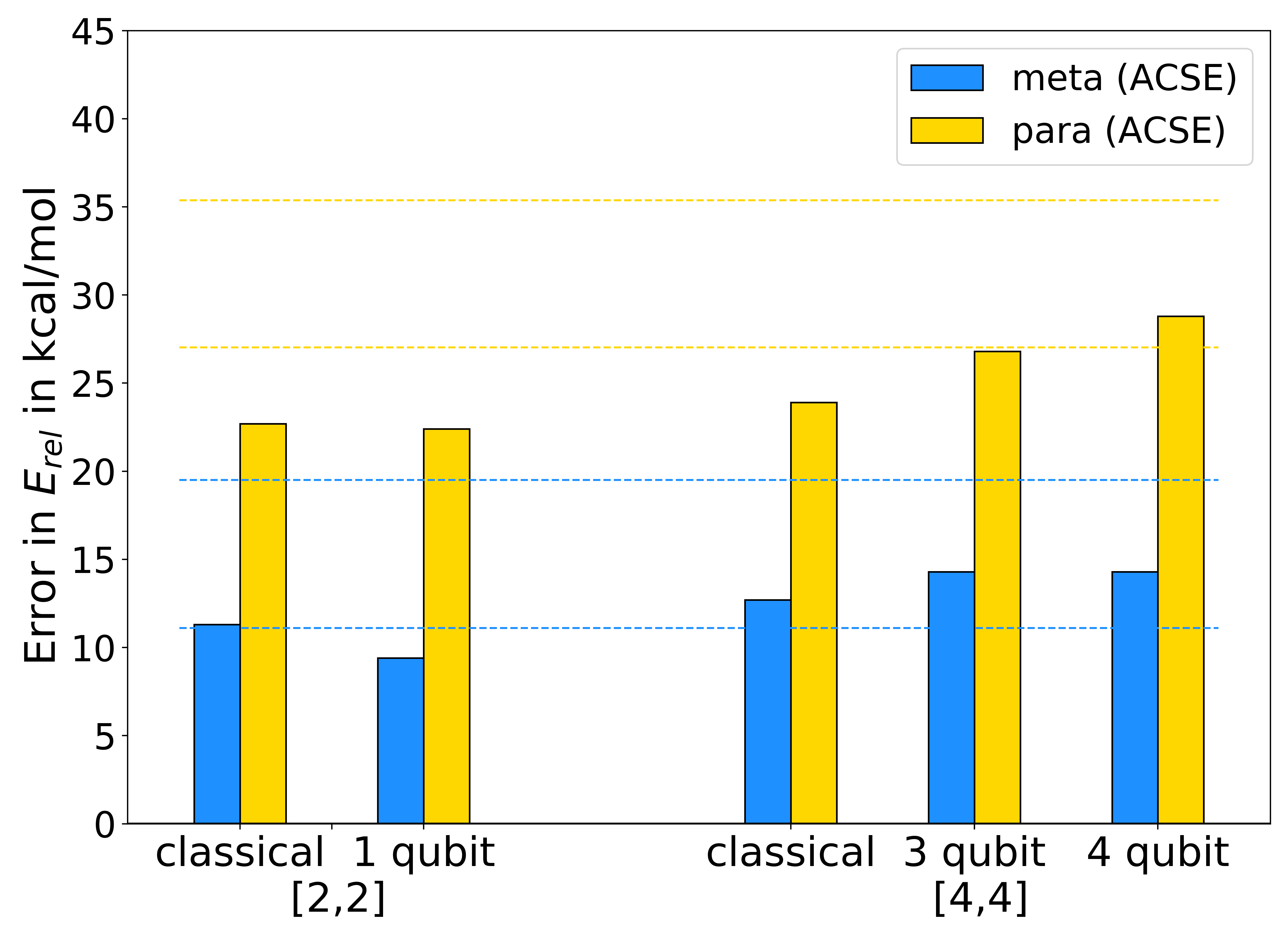}
        \caption{}
        \label{fig:ErelACSE}
        \end{subfigure}
    \begin{subfigure}[t]{0.48\textwidth}
        \centering
        \includegraphics[scale=0.275]{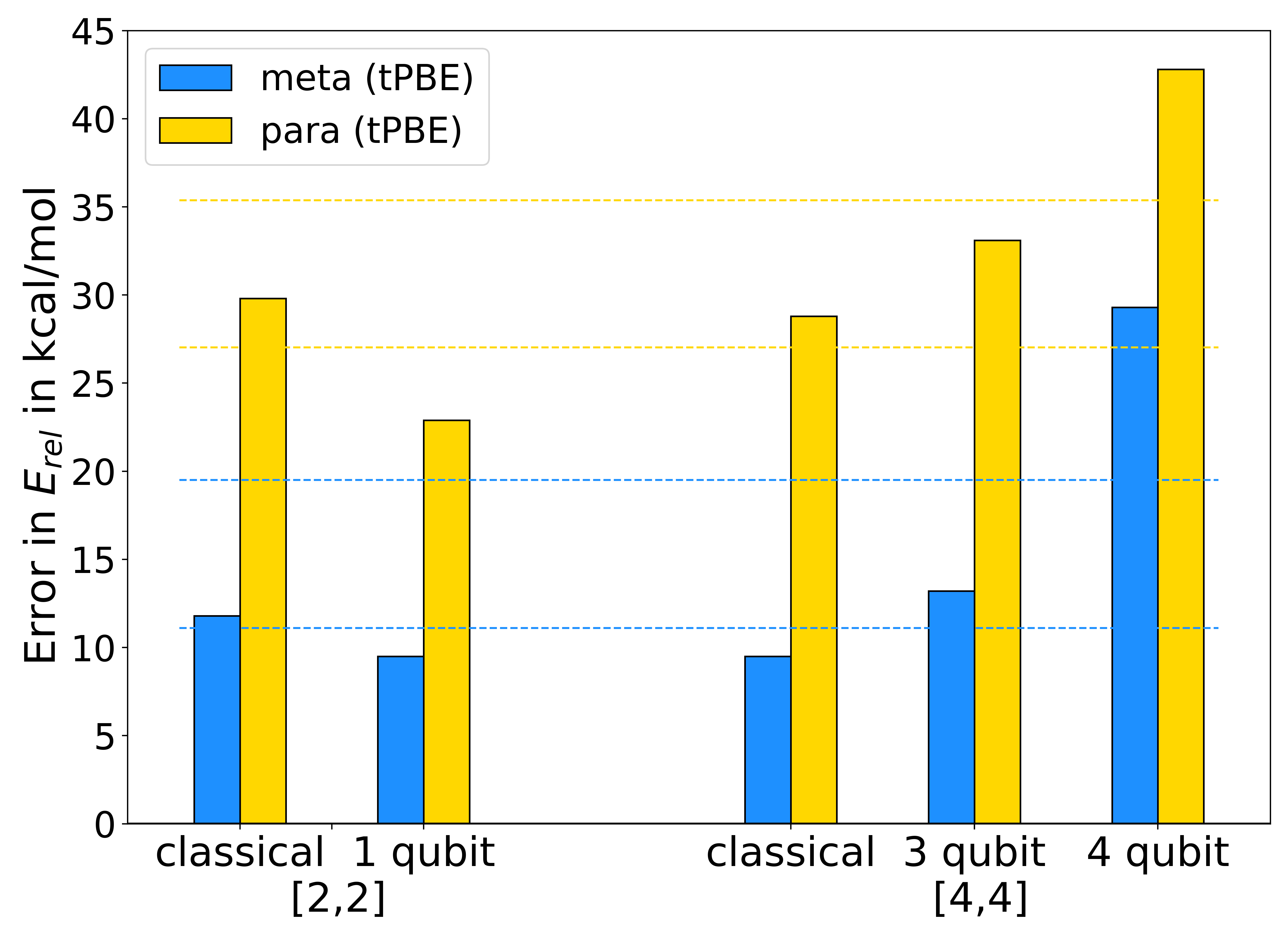}
        \caption{}
        \label{fig:EreltPBE}
    \end{subfigure}
    \medskip
    \begin{subfigure}[t]{.48\textwidth}
        \centering
        \includegraphics[scale=0.275]{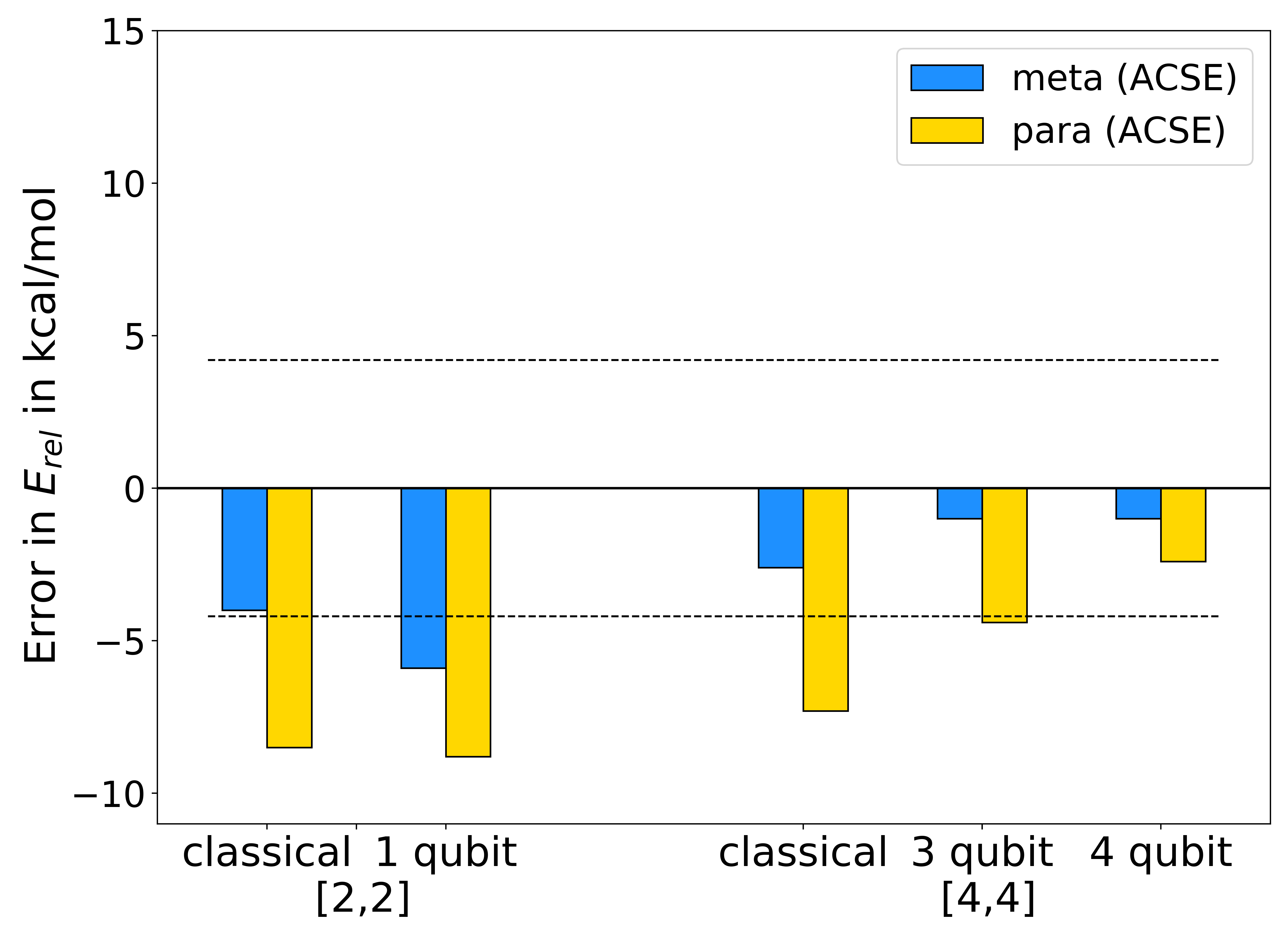}
        \caption{}
        \label{fig:ErrorACSE}
    \end{subfigure}
    \begin{subfigure}[t]{0.48\textwidth}
        \centering
        \includegraphics[scale=0.275]{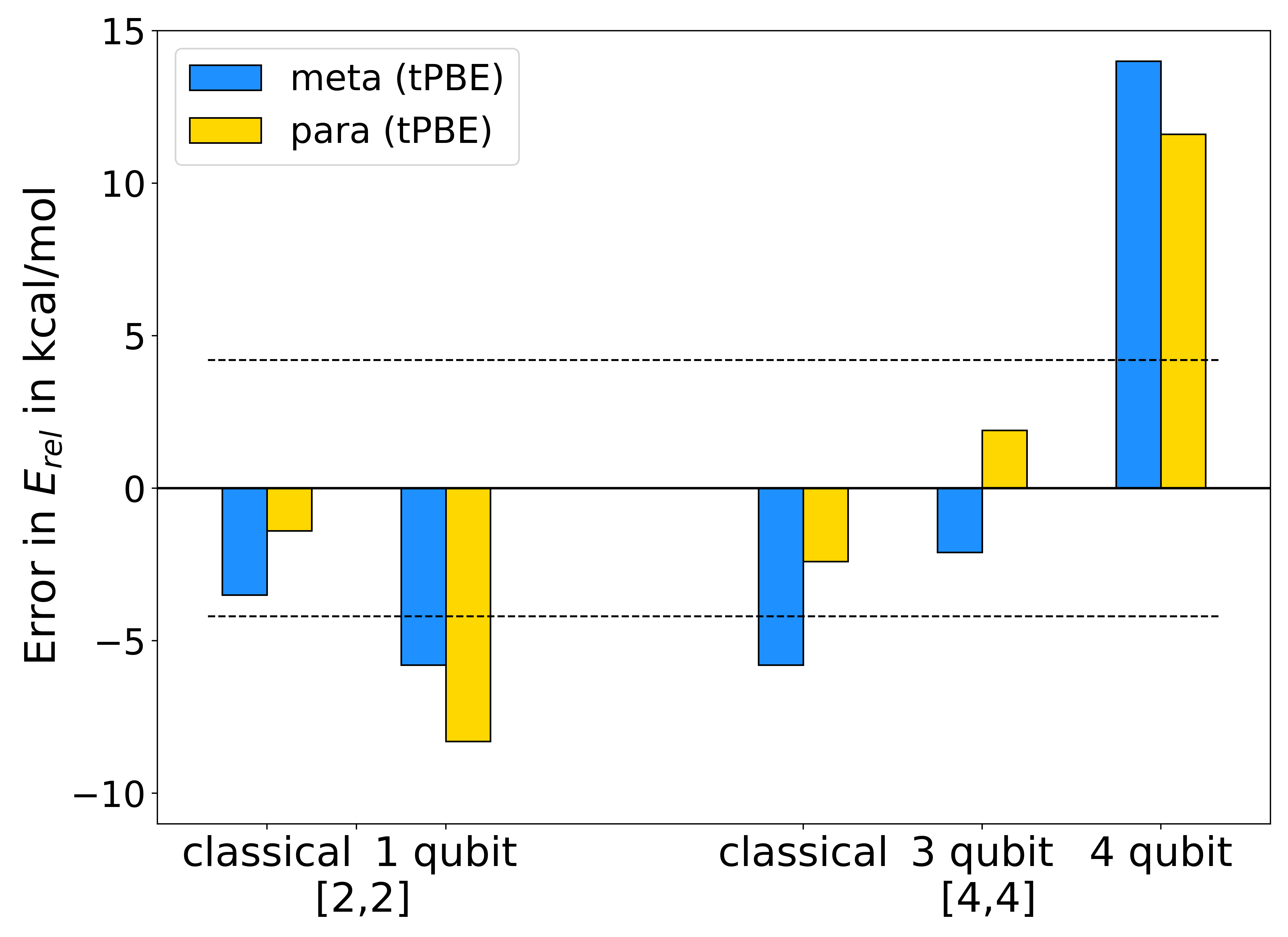}
        \caption{}
        \label{fig:ErrortPBE}
    \end{subfigure}
    \caption{Top row: Relative energies of the meta and para benzynes calculated with ACSE (a) and MC-PDFT (b). Bottom row: Deviations of the relative energies from the corresponding experimental relative energies of the meta and para benzynes calculated with ACSE (c) and MC-PDFT (d). The ortho isomer serves as the reference to determine energies of meta and para benzynes. On the x-axis, "classical" refers to the solutions obtained with a CASSCF 2-RDM evaluated on a classical computer, and "1 qubit", "3 qubit", and "4 qubit" labels indicate ACSE or MC-PDFT (tPBE functional) calculations seeded with a QACSE 2-RDM measured with 1 qubit, 3 qubits, or 4 qubits, respectively. The dashed lines represent uncertainties of the experimental relative energies at the 95\% confidence level.}
    \label{fig:Rels}
\end{figure*}

While the complete active space self consistent field (CASSCF) and configuration interaction (CASCI) methods capture most of the static correlation, they lack a fraction of dynamic correlation resulting in residual errors in the energies of the ortho-, meta-, and para- isomers. Taking the ortho isomer as a reference, classical CASSCF calculations with the correlation-consistent polarized double-zeta (cc-pVDZ) basis\cite{ccpvdz} yield relative energies of 15.2 kcal/mol (meta) and 23.5 kcal/mol (para) in a [2,2] active space, which increase to 16.5 kcal/mol (meta) and 29.5 kcal/mol (para) upon moving to a [4,4] active space. Using electron integrals obtained from the classical CASSCF calculations, the QACSE has been shown to resolve the relative energies of the benzyne isomers with comparable accuracy to classical CASSCF calculations, yielding gaps of 13.8 kcal/mol (meta) and 21.7 kcal/mol (para) in the [2,2] active space performed on a 1 qubit device and 21.3 kcal/mol (meta), 31.0 kcal/mol (para) and 17.6 kcal/mol (meta), 27.8 kcal/mol (para) in a [4,4] active space using 3 and 4 qubit devices, respectively. These results are thoroughly discussed in reference \cite{qACSEBenzyne} and data are shown in Table S1. Having demonstrated the ability to perform accurate CAS calculations on a NISQ device and yielding accurate 2-RDMs with the QACSE method, we take the next step in bringing quantum computing to the realm of applicable quantum chemical computation by uniting the QACSE with the 2-RDM dependent MC-PDFT and ACSE methods to recover all electron correlation in a hybrid quantum-classical approach. \\

The 2-RDMs obtained from the QACSE calculations are used to seed the classical ACSE and various MC-PDFT functionals to calculate the total electronic energy of the three different isomers. The meta-ortho and para-ortho energy gaps, obtained with the ACSE and MC-PDFT (tPBE functional) from CASSCF [2,2] and [4,4] calculations on a classical computer (CASSCF) as well as a quantum computer (QACSE CASSCF) with 1 qubit, 3 qubits, or 4 qubits, are displayed in Figure \ref{fig:ErelACSE} and Figure \ref{fig:EreltPBE}, while Figure \ref{fig:ErrorACSE} and Figure \ref{fig:ErrortPBE} show the respective errors in relative energies as compared to experiment. Using the minimal [2,2] active space required to resolve the biradical character, both the ACSE and MC-PDFT seeded with the QACSE 2-RDMs deliver results that yield a lower bound to the gap obtained using the classical CASSCF 2-RDM. Deviations from the classically seeded references are less than 2 kcal/mol in the case of the ACSE and less than 7 kcal/mol for tPBE, yielding errors slightly outside the bounds of the experimental confidence intervals. \\

A move to the larger [4,4] active space reduces the error of the relative energies in both the underlying classical CASSCF reference and QACSE calculations. Utilizing different symmetry adaptations on the quantum device, the [4,4] QACSE calculations were carried out using either 3 qubits or 4 qubits with the 4 qubits, while producing noisier results, yielding more accurate relative energies. This is carried over in the post-CASSCF ACSE calculation, where favorable error cancellation leads to further increases in the meta-ortho and para-ortho gaps and correspondingly reduced errors as compared to experiment. The meta-ortho gap is reproduced within the experimental confidence intervals for yielding identical gaps of 14.3 kcal/mol both 3 and 4 qubit 2-RDMs, while the para-ortho gap of 26.8 kcal/mo in the 3 qubit case lies barely 0.2 kcal/mol outside the experimental interval and is improved to 28.8 kcal/mol with the use of a 4 qubit 2-RDM. It is noteworthy that the meta-ortho gap is captured with significantly greater accuracy by the ACSE than the para-ortho gap. \\

The trend observed in the ACSE data are reversed in the tPBE calculations and in the classically seeded tPBE calculations the magnitude of relative energies is reduced upon moving from a [2,2] to a [4,4] active space, and the calculated meta-ortho gap lies slightly outside the experimental confidence interval. However, using the 3 qubit QACSE 2-RDM we again observe an increase in the calculated relative energies, which while minor yields relative energies of 13.2 kcal/mol (meta-ortho) and 33.1 kcal/mol (para-ortho) which lie well within the experimental bounds of errors. Using the 2-RDM obtained from 4 qubit NISQ device calculation, the MC-PDFT energies are more prone to errors propagated from noisy 2-RDMs, resulting in significant overestimation of meta-ortho and para-ortho gaps. \\ 
In addition to tPBE we also survey the ftPBE, trevPBE, tBLYP and tPBE0 functionals. The complete data for the meta-ortho and para-ortho energy gaps for all methods are shown in Table S1. The choice of the on-top GGA functional does not significantly affect the predicted relative energies of the three benzyne diradicals in MC-PDFT. The energies obtained with the fully-translated functional ftPBE yield the most accurate results; however, they present only slight improvements upon the tPBE, trevPBE and tBLYP results. The translated hybrid tPBE0 functional is an outlier in the set of surveyed functionals, and it results in large deviations from the experimental energies, particularly in the para-ortho case.  Both the ACSE, as well as MC-PDFT making use of quantum computed active space 2-RDMs deliver relative energies of promising accuracy when compared to both experimental, as well as classical reference data. \\

\subsection{Absolute Energies}
\begin{figure}[t]
    \centering
    \includegraphics[scale=0.275]{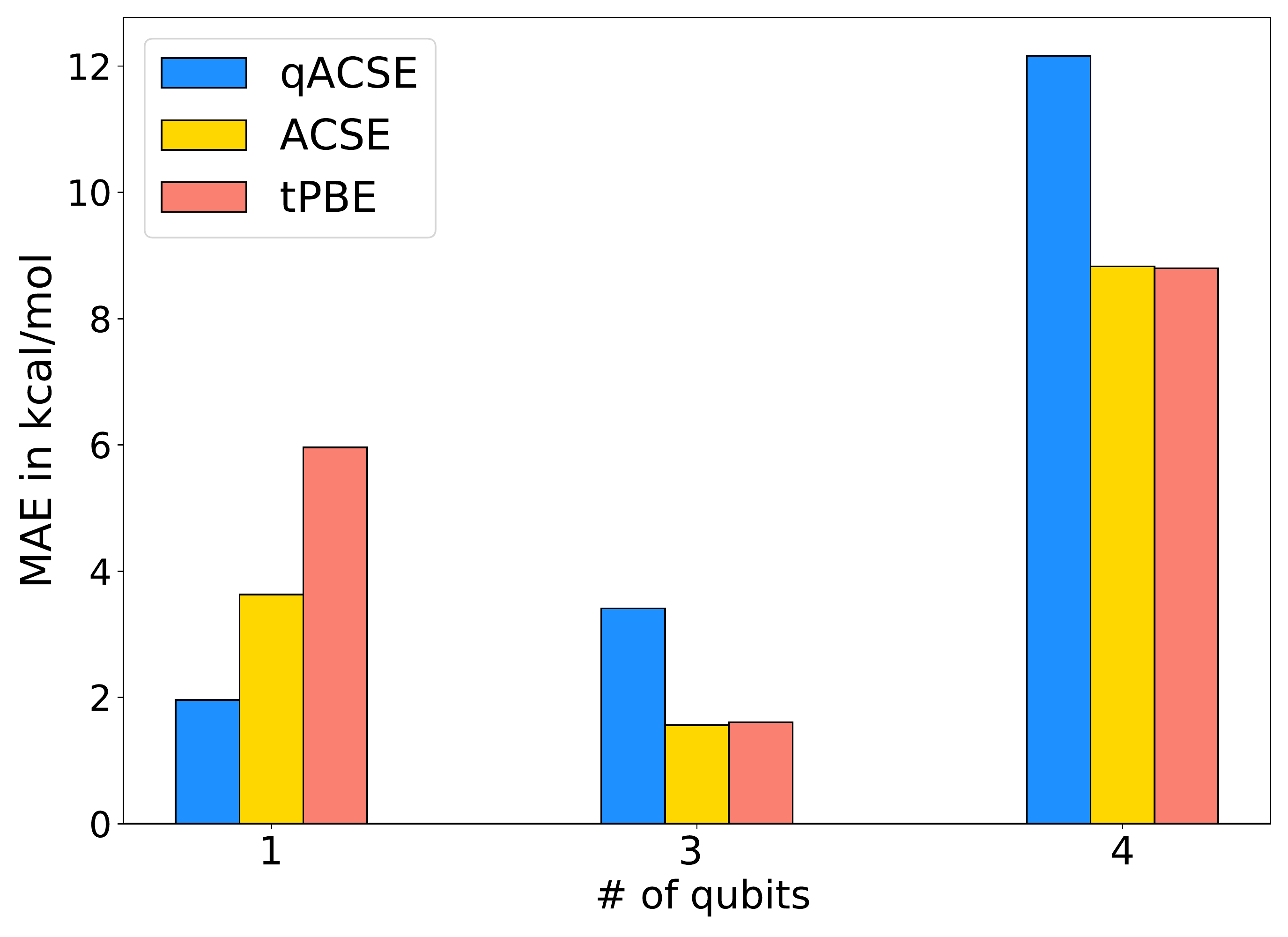}
    \caption{MAEs in kcal/mol of the quantum solution with respect to the classical reference over the three isomers for the QACSE CAS, as well as the QACSE 2-RDM seeded ACSE and tPBE functional MC-PDFT calculations. 1 qubit data uses a [2,2] active space while the 3 and 4 qubit data was obtained with a [4,4] active space.}%
    \label{fig:MAEABS}
\end{figure}

\begin{table}[]
    \centering
    \begin{tabular}{ccc|ccc}
         &  &  & \multicolumn{3}{c}{error in kcal/mol} \\
         &  & & qACSE & ACSE & tPBE \\
        \hline
        [2,2] & 1Q & ortho & 3.03 & 4.39 & 8.99 \\
        & & meta & 1.64 & 2.46 & 6.71  \\
        & & para & 1.21 & 4.04 & 2.18 \\
        \hline
        [4,4] & 3Q & ortho & 1.31 & 0.08 & -3.30  \\
        & & meta & 6.08 & 1.63 & 0.44 \\
        & & para & 2.83 & 2.96 & 1.08 \\
        & 4Q & ortho & 12.36 & 6.63 & -7.52 \\
        & & meta & 13.45 & 8.25 & 12.32 \\
        & & para & 10.67 & 11.59 & 6.57 \\
    \end{tabular}
    \label{tab:errorABS}
    \caption{Deviations in kcal/mol from the classical, or classically seeded calculation, $\Delta E = E_{qc} - E_{c}$, where $E_{qc}$ denotes the qACSE CASCI energy, or the ACSE or tPBE energy when seeded with the qACSE 2-RDM, and $E{c}$ denotes the classical CASSCF energy, or the ACSE or tPBE energy obtained with a CASSCF 2-RDM.}%
\end{table}

To further analyze the impact of the use of NISQ derived QACSE 2-RDMs to seed ACSE and MC-PDFT calculations in a quantum-classical hybrid implementation, we compare the obtained absolute energies to those yielded with a CASSCF seed. Figure 3 shows the mean absolute deviations, defined as $MAE = (\sum_{m,o,p} E_{\text{q}} - E_{\text{c}})/3$, where $\sigma_{m,o,p}$ indicates a sum over the three isomers, and $E_q$ indicates the quantum calculation's energy, and $E_c$ indicates the energy derived from the classical reference calculation, for the QACSE, the QACSE/ACSE and QACSE/tPBE calculations of the total electronic energy for the three benzyne isomers. The QACSE calculation yields energies that in these cases are above those from the classical CASCI calculation in the given molecular orbital basis, and as expected, the magnitude of the deviation increases with increasing qubit count and correspondingly greater noise in the quantum device measurements. The deviations for the absolute energies of the individual isomers can be found in Table \ref{tab:errorABS}. \\

The same behavior is observed in the QACSE/ACSE calculations, which when seeded with quantum computed 2-RDM yield energies above those from the QACSE/ACSE calculations seeded with classical CASSCF. In the [2,2] case, the error from the QACSE with an MAE of 1.96~kcal/mol persists in the ACSE solution with its MAE of 3.63 kcal/mol. In contrast, a reduction in errors is observed in the [4,4] calculations.  Here, while the error obtained by the QACSE/ACSE calculation for the para isomer remains near-identical to that of the underlying CAS calculation, we yield significantly reduced errors in the ortho and meta isomers. The MAE of 1.56 kcal/mol presents a notable reduction compared to the underlying QACSE which displayed a MAE of 3.41 kcal/mol. While the noisier 4 qubit calculation produces larger deviations from the classical calculations, the errors in the QACSE again do not propagate through to the post-CI ACSE calculation and instead the errors in the underlying QACSE calculation are reduced significantly by $\Delta MAE = 3.33$ kcal/mol in the post-CI result.\\

Similar trends are observed using MC-PDFT functionals. Here tPBE yields increased errors over the CAS calculation alone when seeded with the [2,2] QACSE 2-RDM, while reduced errors compared to the CAS calculation alone are obtained in the two studied [4,4] settings, with reductions in MAE of 1.8 kcal/mol and 3.36 kcal/mol in the 3 and 4 qubit cases, respectively. Note that in MC-PDFT errors are lowest in the meta and para isomers, while use of the QACSE solution leads to a large negative deviation from the CASSCF/MC-PDFT solution in the ortho case, where a lower bound to the classical solution is obtained. This is observed in both the 3 qubit and 4 qubit [4,4] calculations, but not in the [2,2] 2 qubit case. \\
Additional insight into the nature of errors introduced via the use of a NISQ deviced 2-RDMs in the ACSE and MC-PDFT calculations may be obtained by decomposing the total electronic energy into its individual 1- and 2-electron components. This work can be found in section 2 of the Supplementary Information. \\

\subsection{Orbital Occupation Numbers}
\begin{table}[]
    \centering
    \begin{tabular}{ccc|cc|cc}
        & \multicolumn{2}{c}{}  & \multicolumn{2}{c}{CAS} & \multicolumn{2}{c}{ACSE} \\
        & \multicolumn{2}{c}{}  & HONO & LUNO & HONO & LUNO \\
        \hline
        [2,2] & ortho & classical & 0.905 & 0.095 & 0.886 & 0.110 \\
                              &  & 1 qubit & 0.845 & 0.155 & 0.833 & 0.163 \\
        & meta & classical & 0.855 & 0.145 & 0.837 & 0.156 \\
                              &  & 1 qubit & 0.800 & 0.200 & 0.787 & 0.206 \\
        & para & classical & 0.615 & 0.385 & 0.608 & 0.386\\
                              &  & 1 qubit & 0.565 & 0.435 & 0.558 & 0.435 \\
        \hline 
        \hline 
        [4,4] & ortho & classical & 0.905 & 0.095 & 0.888 & 0.108  \\
                              & & 3 qubit & 0.925 & 0.075 & 0.904 & 0.092 \\
                              & & 4 qubit & 0.970 & 0.025 & 0.943 & 0.050 \\
        & meta & classical & 0.880 & 0.120 & 0.859 & 0.134 \\
                              & & 3 qubit & 0.860 & 0.120 & 0.840 & 0.153 \\
                              & & 4 qubit & 0.785 & 0.215 & 0.771 & 0.223 \\
        & para & classical & 0.615 & 0.385 & 0.610 & 0.384 \\
                              & & 3 qubit & 0.575 & 0.425 & 0.568 & 0.426 \\
                              & & 4 qubit & 0.600 & 0.395 & 0.593 & 0.396 \\
    \end{tabular}
    \caption{Natural occupation numbers (NON) of the HONO and LUNO orbitals for the CASSCF and QACSE calculations (CAS column) and as well as the QACSE and CASSCF seeded ACSE calculations (ACSE column).}
    \label{tab:NON}
\end{table}

Lastly, we consider the impact of the use of a quantum computed 2-RDM on the natural occupation numbers (NON) obtained from the classical total correlation calculations. As MC-PDFT does not reoptimize the 2-RDM and orbital occupations, we only consider the ACSE results. Table~\ref{tab:NON} shows the NON for the CAS and ACSE calculations, for both classical and quantum computations. Generally, a more correlated solution is characterized by more fractional NON. According to deviations of the NON from 0 and 1 with greater deviations indicating more correlation, the quantum CAS (QACSE) results are more correlated than the classical CASSCF results for all three isomers in the [2,2] active space and for the more strongly correlated meta and para isomers in the [4,4] active space. The use of 4 qubits over 3 qubits yields reduced correlation in ortho and para isomers and increased correlation in the meta isomer. While the dynamic correlation introduced by the post-CI ACSE calculation yields a more correlated solution via the inclusion of core and virtual orbital contributions, the changes in HONO and LUNO occupations that are relevant for the capture of the biradical character displayed by the benzynes are minor compared to the magnitude of the multi-reference character obtained by the QACSE and CASSCF calculations. Consequently, we see a strong dependence on the initial results of the QACSE calculation in the QACSE/ACSE results and as such, computations involving the NISQ device to obtain active space 2-RDMs tend to yield more multi-reference character and more partial HONO and LUNO NON in a post-processing ACSE calculation when compared to the classical reference.  \\

\section{Conclusions}

Realizing the unique position of the ACSE and MC-PDFT as possible post-processing methods to compute all-electron correlation in hybrid quantum-classical algorithms owing to their dependence on only the 2-RDM rather than the $N$-electron wave-function, we have successfully used them in tandem with a CAS calculation, performed with QACSE, to resolve the total electronic correlation energy in the isomers of benzyne. We have demonstrated that 2-RDMs from NISQ devices, after their error mitigation with necessary $N$-representability conditions, allow for the resolution of an experimentally verifiable quantity, the relative energies of the different benzyne isomers, within the bounds of the experimental margins of error in a QACSE/ACSE and QACSE/MC-PDFT hybrid classical-quantum algorithm. Furthermore, we have shown that the errors arising in CAS calculations on NISQ devices are not amplified in post-correction ACSE and MC-PDFT calculations relying on their 2-RDMs. Instead in the noisier 3 and 4 qubit calculations we observe a reduction of the error compared to the respective classical analogues when comparing the absolute energies of the CAS and post-CI calculations. Thus classical post-correction calculations of the total correlation energy, as in  QACSE/ACSE and QACSE/MC-PDFT,  may play an additional role as further sources of error correction in the applications of quantum algorithms. \\

With the advances reported in this article we successfully take a first step in bringing NISQ device based hybrid quantum-classical algorithms into the realm of everyday computational chemistry applications.  Even as quantum hardware markedly improves, the quantum-classical quantum algorithm presented here, including the 2-RDM-based error mitigation, will be critically important for merging the strengths of quantum and classical computers for accurately simulating the energies and properties of chemically important molecules and materials\cite{AspuruGuzikRev, ChanCanonicalT, KadeRev, Reiher_QCRxnMech, vonburg2021quantum}.

\section{Methods}
\subsection{ACSE}
The $N$-electron Schr\"odinger equation may be contracted onto the space of pairwise excitations, giving rise to the contracted Schr\"odinger equation (CSE)\cite{CSE}: 
\begin{equation}
    \bra{\Psi}\hat{a}^{\dagger}_i \hat{a}^{\dagger}_j \hat{a}_l  \hat{a}_k  \hat{H}\ket{\Psi} = E\; {}^2D^{i,j}_{k,l} \,,
\end{equation}
where $\hat{a}^{\dagger}$ and $\hat{a}$ are the fermionic creation and annihilation operators, respectively, and $\hat{H}$ is the electronic Hamiltonian:
\begin{equation}
    \hat{H} = \sum_{ij} {}^1K^i_j \hat{a}^{\dagger}_i \hat{a}_j + \sum_{ijkl} {}^2V^{ij}_{kl} \hat{a}^{\dagger}_i \hat{a}^{\dagger}_j \hat{a}_l \hat{a}_k \,,
\end{equation}.
The CSE may be expanded into its Hermitian and anti-Hermitian parts, 
\begin{equation}
    \bra{\Psi}\{\hat{a}^{\dagger}_i \hat{a}^{\dagger}_j \hat{a}_l  \hat{a}_k,(\hat{H}-E)\}\ket{\Psi} + \bra{\Psi}[\hat{a}^{\dagger}_i \hat{a}^{\dagger}_j \hat{a}_l  \hat{a}_k,(\hat{H}-E)]\ket{\Psi} = 0 \,,
\end{equation}
where the square and curly brackets denote the commutator and anti-commutator, respectively. While the CSE depends on the 2-, 3-, and 4-reduced density matrix (RDM), we can select just the anti-Hermitian part of the CSE, yielding the anti-Hermitian contracted Schr\"odinger equation (ACSE), which in turn only depends on the 2-, and 3-RDMs\cite{RDMbook, CSE, ACSE, ACSE2, ACSE3, ACSE4}: 
\begin{equation}
    \bra{\Psi}[\hat{a}^{\dagger}_i \hat{a}^{\dagger}_j \hat{a}_l  \hat{a}_k,\hat{H}]\ket{\Psi} = 0 \,.
\end{equation}
The computationally expensive dependence on the 3-RDM may be further simplified by employing cumulant reconstruction, allowing the 3-RDM elements to be approximately reconstructed in terms of the 2-RDM elements\cite{CSE,Schwinger, 35CSE}:
\begin{equation}
    {}^3D^{ijk}_{qst} \approx {}^1D^{i}_{q} \wedge {}^1D^{j}_{s} \wedge {}^1D^{k}_{t} + 3 {}^2\Delta^{ij}_{qs} \wedge {}^1D^{k}_{t} \,,
\end{equation}
where
\begin{equation}
    {}^2\Delta^{ij}_{qs} = {}^2D^{ij}_{qs} - {}^1D^{i}_{q} \wedge {}^1D^{j}_{s} \,,
\end{equation}
and $\wedge$ denotes the antisymmetric Grassmann wedge product\cite{CSE}, which is defined as:
\begin{equation}
    {}^1D^{i}_{k} \wedge {}^1D^{j}_{l} = \frac{1}{2}({}^1D^{i}_{k}{}^1D^{j}_{l} - {}^1D^{i}_{l} {}^1D^{j}_{k}) \,.
\end{equation}
In this implementation we use the Valdemoro reconstruction, further simplifying our 3-RDM reconstruction by approximating ${}^3\Delta^{ijk}_{qst}$ to be zero\cite{cumulant, Valdemoro}. \\

The ACSE is solved via a series of differential equations, where at each step we use the solution of the ACSE to generate an anti-Hermitian operator, $A$, which is related to the gradient information with respect to various two-body body unitary transformations. Indeed, the exponential of the operator gives a unitary transformation which allows us to minimize the energy against these transformations\cite{ACSE2}. If the starting 2-RDM corresponds with the single-reference state, e.g. obtained from a Hartree-Fock calculation, the obtained solutions resolve dynamic correlation effects with accuracy comparable to CCSD(T)\cite{ACSECC, ACSElb}. However, by starting with a multi-reference 2-RDM, which can be obtained from active space methods, the total electron correlation is obtained---not just the static correlation present in a multiconfigurational seed 2-RDM\cite{ACSE3, ACSE2, ACSEMR}. This approach has been demonstrated to resolve successfully the chemical and electronic properties in a variety of correlated systems, such as the S-T gaps in small biradicals\cite{ACSE3}, the barriers and conical intersections in chemical reactions\cite{ACSErxn1, ACSErxn2, ACSERxn3}, and excited states\cite{ACSEES2, ACSEES} with accuracy exceeding go-to methods such as MRCI+Q or CASPT2, and comparable to state-of-the art methods such as AFQMC. \\

\subsection{QACSE}

The quantum ACSE (QACSE) method relies upon the capacity of a quantum computer to efficiently simulate quantum states in potentially two regards. First, the exponential of the $A$ matrix can be implemented efficiently through methods of quantum computation, and then by performing reduced state tomography, we can measure the 2-RDM of a prepared quantum state at any given iteration. The tomography naively requires $O(r^4)$ measurements, although this can be reduced to $O(r^2)$ through a variety of techniques \cite{tomography1, tomography2}. \\

Second, elements of the $A$ matrix can be obtained efficiently through the preparation of an auxiliary state involving a single propagator step. Tomography of the $A$ matrix scales with the order of the $A$ operator, which generally will be $O(r^4)$.  In both cases, the use of a quantum computer allows us to extract information directly from the quantum state, circumventing the cost and errors which come from reconstruction\cite{qACSE,qACSEBenzyne}.

\subsection{Quantum Computation}

The 2-RDMs used in the present work were obtained by using quantum devices (ibmq\_armonk for the 1-qubit results, and ibmq\_bogota for the 3- and 4-qubit results) through the IBM Quantum Experience\cite{IBMQuantum}. We use the \textsc{python} 3 package \textsc{qiskit} (v 0.15.0)~\cite{Qiskit} to interface with the device. The calculations themselves are multifaceted, with significant error mitigation being required. Full details regarding these calculations can be found in Ref. \cite{qACSEBenzyne}.

For this work we obtained the 2-RDMs on the quantum computer, and calculated the ${}^2 A$  matrix classically with a reconstructed 3-RDM. For all of these calculations, the error due to reconstruction in each active space was less than 1 millihartree, and was significantly less than errors introduced due to noise on the quantum computer. \\

\subsection{PDFT}

The multiconfiguration pair-density functional theory (MC-PDFT) is an extension of the density functional theory to the molecular systems with strong electron correlation. The advantage of MC-PDFT is that the correlation energy is computed  with a low computational cost compared to the alternative multireference methods, such as multireference second order perturbation theory (MRPT2)\cite{MCPT} or multireference configuration interaction (MRCI)\cite{MRCI}. In MC-PDFT, kinetic and classical Coulomb energies are evaluated from the multiconfigurational wave function represented by 1,2-RDMs, while the rest of the energy is  computed non-iteratively as a functional of the total electron density, the on-top pair density, and their spatial derivatives\cite{MCPDFT1}:

\begin{equation}
\begin{split}
    E_{\text{MC-PDFT}} = & V_{nn} + \sum_{pq} h^p_q D^p_q + \frac{1}{2}\sum_{pqrs} g^{pr}_{qs} D^p_q D^r_s   \\
    & + E_{\text{ot}}[\rho(D^p_q), \rho^{'}(D^p_q), \Pi(d^{pr}_{qs}), \Pi^{'}(d^{pr}_{qs})] \,.
\end{split}
\end{equation}

While the total electron density, $\rho$, defines the probability of locating a single electron at the given point of space, the on-top pair density, $\Pi$, describes the probability of observing two electrons of the opposite spin at the same space point. The dependence of the functional upon these densities conveniently evaluated from 1,2-RDMs resolves the symmetry dilemma\cite{SymmetryDilemma} and ensures the correct asymptotic behavior of the total energy at the dissociation limit. In general, any type of multiconfigurational reference can be exploited in MC-PDFT. The most common choices are state-specific and state-averaged complete active space self-consistent field (SS-CASSCF and SA-CASSCF) wave functions\cite{choices1, choices2, choices3,choices4}, however, other multiconfigurational wave functions have also been successfully combined with MC-PDFT including those obtained with restricted active space self-consistent field (RASSCF)\cite{RASSCF1,RASSCF2} restricted active space configuration interaction (RASCI)\cite{RASCI1, RASCI2, RASCI3},generalized active space self-consistent field (GASSCF)\cite{GASSCF1, GASSCF2, GASSCF3, GASSCF4}, and density matrix renormalization group (DMRG)\cite{DMRG1, DMRG2, DMRG3} methods. In addition to that, the wave function framework can be completely abandoned in MC-PDFT by using the variational 2-RDM method, where the 2-RDM is computed directly under constraint of N-representability conditions\cite{V2PDFT}. All these approximations aim to reduce the computational cost that scales exponentially with the system size and reach a polynomial scaling so that the larger number of electrons can be correlated, which is important in strongly correlated systems.
Here, we explored another appealing opportunity that is combining MC-PDFT calculations with 2-RDMs simulations performed on the quantum computer. The 1-RDMs were contracted from the qACSE 2-RDMs and the PDFT step was performed on the classical computer. To estimate the sensitivity of predicted ground state energies to the choice of the on-top functional, we explored several types of functionals including: (i) translated\cite{MCPDFT1} (tPBE and tBLYP), (ii) fully-translated\cite{ftPBE} (ftPBE), and (iii) translated hybrid\cite{tPBE0} (tPBE0) density functionals. Note that translation of Kohn-Sham density functionals into the on-top density functionals does not alter the functional form but redefines the spin-densities and their gradients in terms of the total electron density and the on-top pair density.

\begin{acknowledgement}
D.A.M. gratefully acknowledges funding from the U.S. Department of Energy (Office of Basic Energy Sciences) under Award No. DE-SC0019215
and the U.S. National Science Foundation under Award No. CHE-2035876 and No. CHE-1565638. A.O.L. and L.G. acknowledge funding from the Division of Chemical Sciences, Geosciences, and Biosciences, Office of Basic Energy Sciences of the U.S. Department of Energy through Grant DE-SC002183. We acknowledge the use of IBM Quantum services for this work. The views expressed are those of the authors, and do not reflect the official policy or position of IBM or the IBM Quantum team. 
\end{acknowledgement}

\bibliographystyle{chem-acs}
\bibliography{citations.bib}

\end{document}